\documentclass[article,showpacs,superscriptaddress,groupedaddress]{revtex4} 
\usepackage{graphicx}  
\usepackage{amsmath}
\usepackage{amssymb}
\usepackage{dcolumn}  
\usepackage{bm}       
\usepackage{amssymb}

\begin{document}

% the following line is for submission, including submission to the arXiv!!
%\hspace{5.2in} \mbox{Fermilab-Pub-04/xxx-E}

\title{Microscopic Model for the Scintillation-Light Generation and Light-Quenching in CaWO$_4$ single crystals}
\author{S. Roth}
 \email{sabine.roth@ph.tum.de}
\affiliation{Physik-Department, Technische Universit\"at M\"unchen, James-Franck-Stra\ss e 1, D-85748 Garching, Germany}
\author{F.v. Feilitzsch}
\affiliation{Physik-Department, Technische Universit\"at M\"unchen, James-Franck-Stra\ss e 1, D-85748 Garching, Germany}
\author{J.-C. Lanfranchi}
\affiliation{Physik-Department, Technische Universit\"at M\"unchen, James-Franck-Stra\ss e 1, D-85748 Garching, Germany}
\author{W. Potzel}
\affiliation{Physik-Department, Technische Universit\"at M\"unchen, James-Franck-Stra\ss e 1, D-85748 Garching, Germany}
\author{S. Sch\"onert}
\affiliation{Physik-Department, Technische Universit\"at M\"unchen, James-Franck-Stra\ss e 1, D-85748 Garching, Germany}
\author{A. Ulrich}
\affiliation{Physik-Department, Technische Universit\"at M\"unchen, James-Franck-Stra\ss e 1, D-85748 Garching, Germany}

\date{\today}

\begin{abstract}
Scintillators are employed for particle detection and identification using light-pulse shapes and light quenching factors. We developed a comprehensive model describing the light generation and quenching in CaWO$_4$ single crystals used for direct dark matter search. All observed particle-dependent light-emission characteristics can be explained quantitatively, light-quenching factors and light-pulse shapes are calculated on a microscopic basis. This model can be extended to other scintillators such as inorganic crystal scintillators, liquid noble gases or organic liquid scintillators.
\end{abstract}
%===============================================================================================================
\maketitle
\begin{figure}[!h]
 \begin{center}
  \includegraphics[width=0.35\textwidth]{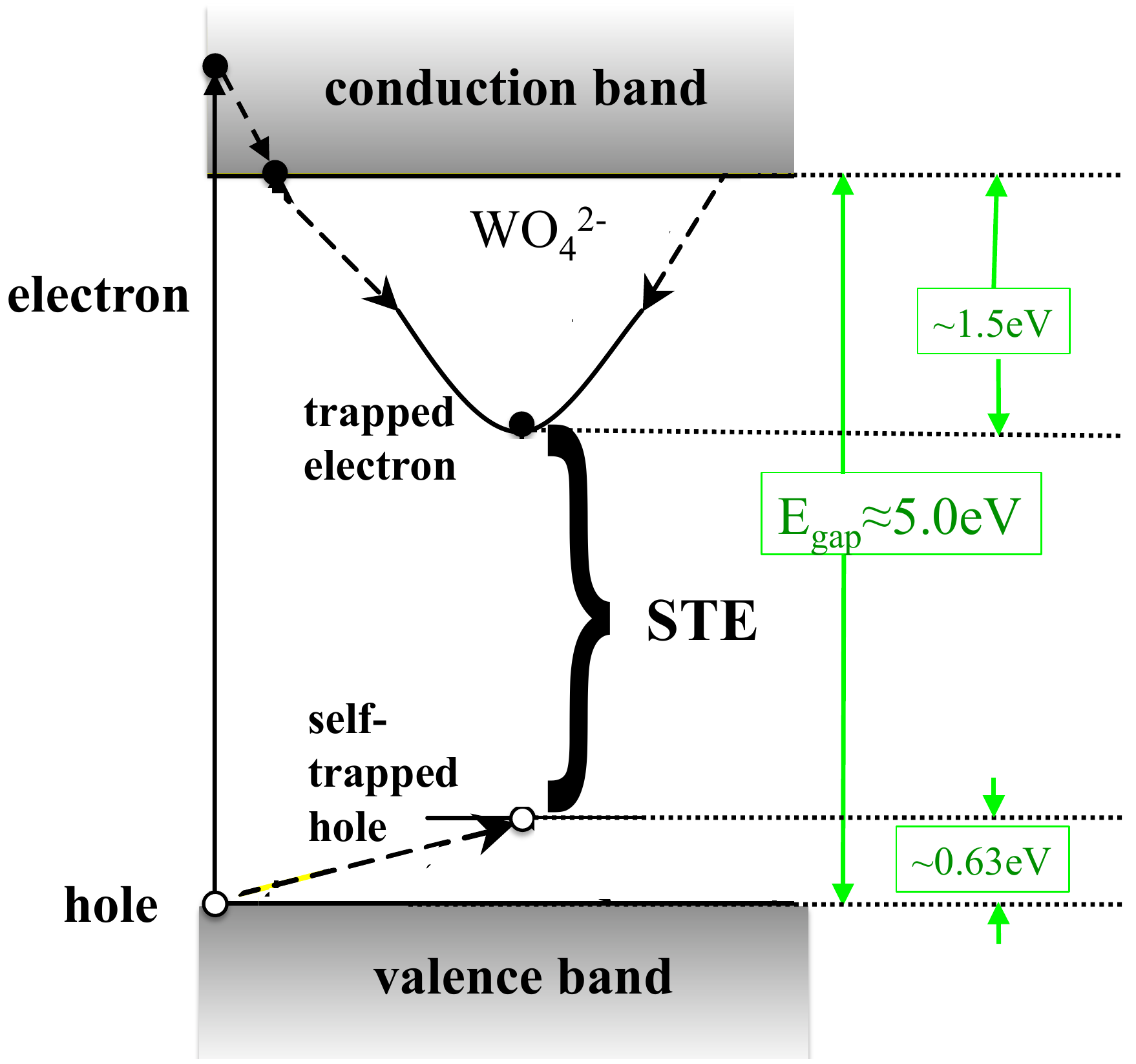}
  \caption{Sketch of the Self-Trapped-Excition (STE) formation process in CaWO$_4$. For details, see main text.}\label{STEformation}
 \end{center}
\end{figure}

%============================================================================================================
Scintillating materials are employed in many different fields of physics for the detection of ionizing radiation and for particle identification. Pertinent examples are rare event searches using crystal scintillators as target material of cryogenic detectors as in the CRESST (Cryogenic Rare Event Search with Superconducting Thermometers \cite{CRESST, CRESSTII}) experiment or liquid noble gases, e.g., liquid Xenon in the Xenon100 \cite{Xenon100} and LUX \cite{LUX} experiments. In the CRESST experiment, different particles interacting in the detector are identified by the so-called light-quenching effect, i.e., different particles depositing the same amount of energy in the detector lead to the generation of different amounts of scintillation light as, e.g., phenomenologically described by Birks' saturation law \cite{BirksI, BirksII}.\\
\newline
However, to quantitatively explain the observed particle-dependent light-emission characteristics we have developed a comprehensive microscopic model describing the light generation and quenching in CaWO$_4$ single crystals for arbitrary interacting particles, i.e., also for nuclear recoils. Within the model the light emission is described in terms of analytical rate equations for the time and position-dependent STE-density evolution at intrinsic scintillation centers and at defective unit cells \footnote{Those are assumed to also be able to produce scintillation light, however, of course, with different efficiencies, rate constants and temperature-dependencies compared to intrinsic centers.}, for details see \cite{RothPhD}.\\
\newline
%============================================================================================================
The model consists of three major parts:
\begin{enumerate}
\item The description of the light-generation mechanism in CaWO$_4$, i.e., the different excitation and de-excitation processes of so-called Self-Trapped Excitons (STEs).
\item The determination and geometrical modeling of the energy-deposition process by a primary interacting particle.
\item The mathematical description of the temporal evolution of this initially created STE population.
\end{enumerate}
\textit{1. Light-Generation Mechanism}\\
CaWO$_4$ is a self-activated intrinsic scintillator where the light emission is well described by the radiative recombination of (Frenckel-type) Self-Trapped-Excitons (STEs) at regular [WO$_4$]$^{2-}$ complexes \cite{RothPhD}. Figure \ref{STEformation} depicts the STE formation process: The deposition of energy in the crystal can lead to the excitation of electrons from the valence band to the conduction band (band-gap energy $E_{gap}\approx 5.0$eV). The holes created in the valence band relax to the band edge, become localized at a [WO$_4^\star$]$^{2-}$ complex \footnote{Asterisk: excited state of the [WO$_4$]$^{2-}$ complex.} where they form Self-Trapped-Holes (STHs) due to a trigonal Jahn-Teller distortion of the electronic valence-band structure. The presence of the STH induces a distortion of the electronic structure of the conduction band at the [WO$_4^\star$]$^{2-}$ complex such that a potential well for an electron is formed. Hence, when an electron relaxes to the band edge (from which direct recombination would be spin-forbidden), it further relaxes into the potential well leading to the creation of a STE with a Jahn-Teller energy of the STH of $\sim0.63$eV \cite{HolesinCaWO4} and an electron trap depth of $\sim1.5$eV \cite{RothPhD, TransientOpticalAbsorptioninCaWO4, BandStrucutreCaWO4}. The possibility of the electron being captured in an electron trap (at a defective [WO$_4$]$^{2-}$ center) and, thus, being removed from the STE-formation process is included in the model. Such trapping events give rise to a limited and crystal-dependent efficiency of the STE-production process and, thus, of the crystal's light yield. The STE formation process is modeled as diffusion-controlled recombination process according to Onsager theory \cite{FromLumNonLin,OnsagerRadius} (see, \cite{RothPhD} for details).

\begin{figure}[!h]
 \begin{center}
  \includegraphics[width=0.46\textwidth]{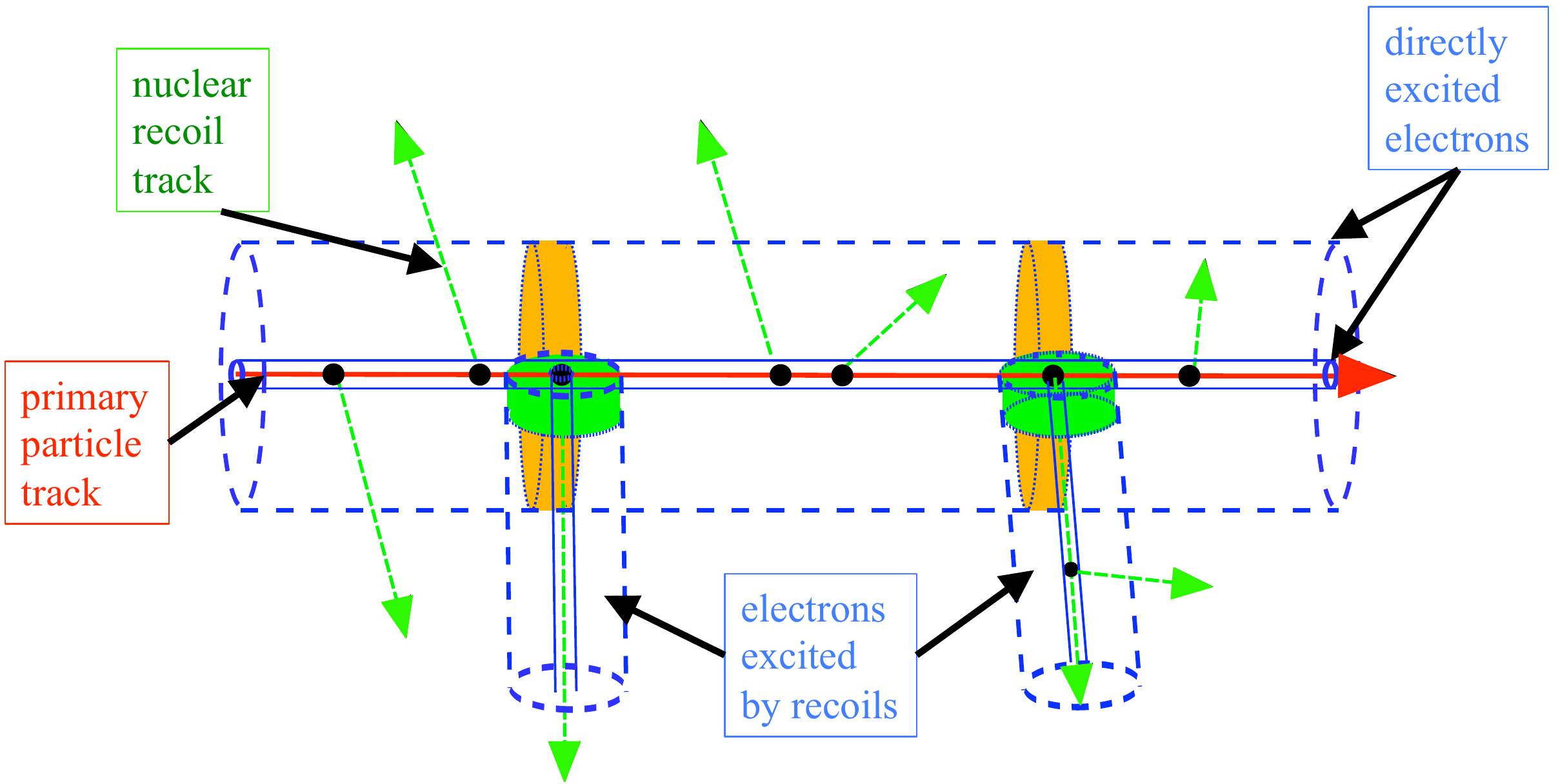}
  \caption{Geometrical model developed for the ionization distribution produced by a heavy, primary interacting particle, such as a recoiling nucleus. For more details, see main text.}\label{GeometricalModelIonizNucleus}
 \end{center}
\end{figure}

\begin{figure}[!h]
 \begin{center}
  \includegraphics[width=0.48\textwidth]{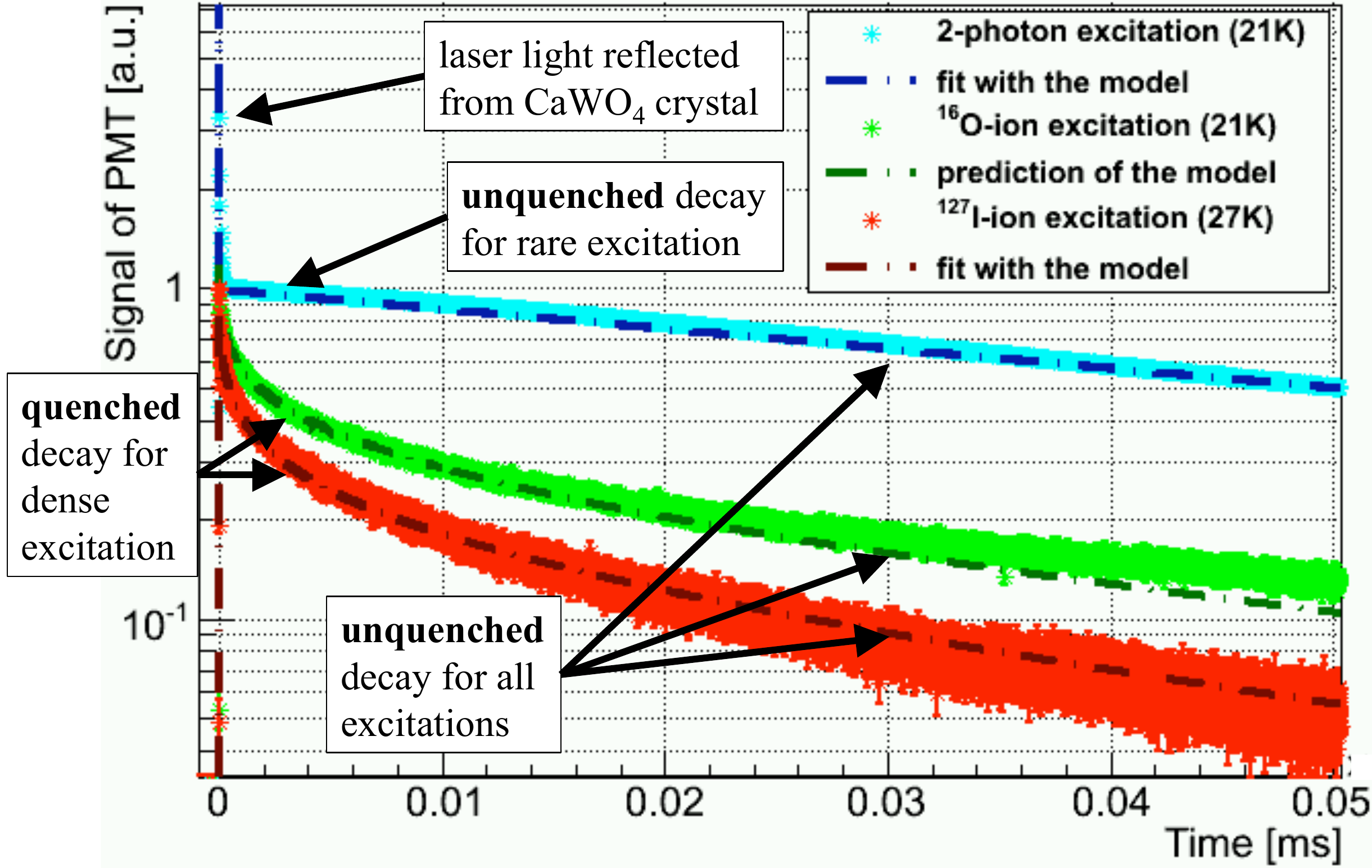}
  \caption{Experimentally recorded decay-time spectra of the scintillation light of a CaWO$_4$ crystal at $\sim 20$K for 2-photon excitation with the N$_2$-laser (blue markers), oxygen ions (green markers) and iodine ions (red markers). As indicated in the figure, the pulse shapes differ according to the excitation mode as predicted by the model: The more closely spaced the excited STE population, the more quenched the scintillation-light generation, i.e., the more distinct the strong decay at the beginning of the pulse. Dashed lines correspond to fits (for laser and iodine-ion excitation) and predictions (for oxygen-ion excitation) of the model.}\label{ExpDTScold500nm}
 \end{center}
\end{figure}

\noindent Within the model, several de-excitation possibilities for the produced STEs are included. STEs can:
\begin{itemize}
\item recombine radiatively, leading to the production of scintillation light
\item recombine non-radiatively, leading to the production of phonons
\item migrate to neighboring regular or defective [WO$_4$]$^{2-}$ complexes
\item interact with each other, by the so-called F\"orster interaction \cite{Foerster}. The strength of this interaction is described by the so-called F\"orster radius $R_{dd}$. The probability for the F\"orster interaction to occur strongly depends on the density of the STEs. Thus, the non-radiative destruction of STEs via the F\"orster interaction, i.e., the \textit{light-quenching effect}, strongly depends on the ionization-density produced by an interacting particle in its energy loss process.
\end{itemize}
Temperature-dependencies (e.g., the temperature dependence of the F\"orster radius) as well as trapping of electrons at defects are included in the model.\\
\newline
\textit{2. Spatial Distribution of the Particle-Induced STE-Production}\\
To calculate the light-emission, the spatial STE-density distribution initially created in the energy deposition of an interacting particle has to be determined taking the F\"orster interaction into account. For this purpose, the ionization-density distribution initially created in the energy-loss process of an interacting particle is simulated (using SRIM \cite{SRIM} for nuclear recoils and CASINO \cite{CASINO} for electron recoils) and described by a geometrical model. Figure \ref{GeometricalModelIonizNucleus} depicts the basic structure of the geometrical model outlining the ionization distribution produced by a heavy, primary interacting particle.
The central red arrow corresponds to the track of the primary particle. Indicated as blue cylinders is the possibility that the primary particle directly excites electrons in its energy-loss process where the energy distribution of the created electrons follows a Landau distribution: The major part of the excited electrons is very low in energy and is therefore situated in the direct vicinity of the primary particle track (inner cylinder, solid blue line). Only a few high-energy electrons are created which distribute their energy, i.e., create further ionization, in a considerably larger volume around the primary particle track (outer cylinder, dashed blue line). Additionally, the possibility is included that the primary particle produces nuclear recoils (black dots with dashed green arrows indicating the initial direction of the produced nuclear recoil). Of course, those recoiling nuclei can themselves produce electrons directly (high- and low-energetic ones, exemplarily indicated by concentric blue cylinders for two nuclear recoils) as well as further nuclear recoils (again indicated by black dots and dashed green arrows as shown for one example). The electron populations excited by the primary particle and by its nuclear recoils overlap in space at the production points of the nuclear recoils. This is included in the model and is shown in figure \ref{GeometricalModelIonizNucleus} as overlap of the orange and green cylinders. With the help of the simulations the particle- and energy-dependent spatial ionization-density distribution within the different volumes is determined (for details see \cite{RothPhD}). The results of the simulations include the effect described by the Lindhard theory \cite{Lindhard}, i.e., the particle and energy dependence of the ionization yield. In the modeling of the spatial distribution of the created ionization additionally, the theories from Bethe-Bloch \cite{Bethe} and Landau \cite{SecondaryEleLandau} are additionally considered (see \cite{RothPhD} for details).\\
\newline
\textit{3. Temporal Evolution of the STE population and Light Generation}\\
Using the STE excitation and de-excitation processes as well as their temperature dependencies (see \textit{part 1} of the model), the temporal evolution of the STE population initially created in the energy-loss process of an interacting particle can be described by a system of coupled non-homogeneous differential equations \cite{RothPhD} for the time- and position-dependent STE density $n_{STE}(\textbf{x},t)$. The spatial integration over $n_{STE}(\textbf{x},t)$, as well as the calculation of the fraction of the STE population decaying radiatively (with radiative decay time $\tau_{rad}(T)$) and escaping the crystal (i.e., not being re-absorbed) leads to an expression for the time-dependent photon-production process: The decay-time spectrum of the created scintillation light. The particle-dependent pulse shape of the scintillation light as predicted by the model can be described by a fast non-exponential decay at the beginning of the pulse, caused by the fast non-radiative destruction of a dense population of STEs  directly after excitation. This fast decay at the beginning of the pulse is followed by a purely exponential decay with a decay time corresponding to the temperature- and crystal-dependent intrinsic lifetime of STEs. This decay time does not depend on the excitation mode, i.e., the initially produced ionization-density distribution, as at this point in the light-production process, the STEs are no longer spaced closely enough to interact with each other. By integrating the decay-time spectrum over time, the particle-, energy-, temperature- and crystal-quality-dependent amount of scintillation light created, can be calculated. For details on the differential equations as well as the mathematical expressions obtained for the decay-time spectrum and the light yield, see reference \cite{RothPhD}.\\
\newline
%============================================================================================================
The expressions for the scintillation-light decay-time spectrum and for the light yield depend on 18 free parameters of the model, e.g., the temperature-dependent radiative decay-time $\tau_{rad}(T)$ or the temperature-dependent F\"orster radius $R_{dd}(T)$. To determine the model parameters, experiments at room temperature and at $\sim 20$K were carried out where CaWO$_4$ single crystals were excited in several ways leading to different STE densities: Pulsed ion-beam excitation (oxygen- and iodine-ion beams at the Maier-Leibnitz-Laboratory in Garching, Germany) leading to dense, particle-induced initial STE populations, and the excitation of CaWO$_4$ by a 2-photon absorption process (pulsed light from a N$_2$-gas laser, $E_{laser-photon}=3.68$eV) generating spatially separated, rare excitations in the crystal. According to the model, a 2-photon excitation should lead to one slow, temperature-dependent exponential decay time of the produced scintillation light, corresponding to the intrinsic lifetime of the STEs, preceded by a temperature dependent, much shorter rise time caused by the STE production process discussed in detail in \cite{RothPhD}. In figure \ref{ExpDTScold500nm}, examples for the decay-time spectra recorded at $\sim 20$K for 2-photon excitation with the N$_2$-laser \footnote{The fast spike visible at the beginning of the scintillation-light pulse excited by pulsed laser excitation is due to laser light being partially reflected from the CaWO$_4$ crystal and does not correspond to the excited CaWO$_4$ scintillation light. For a detailed discussion, see \cite{RothPhD}.} (blue markers), oxygen ions (green markers) and iodine ions (red markers) can be seen (all decay-time spectra were normalized to unity at the beginning of the pulses). Error bars correspond to statistical noise on the PMT baseline and include uncertainties due to systematic drifts of the PMT baseline. In addition to the data, for each decay-time spectrum, the fit of the model to the data (for laser and iodine-ion excitation, used for the determination of the model parameters) or the model prediction (for oxygen-ion excitation, calculated decay-time spectrum with all parameters of the model already fixed) of the light-pulse shape, are shown as (respective) darker dashed line.

\begin{figure*}[t!]
 \begin{center}
  \includegraphics[width=0.95\textwidth]{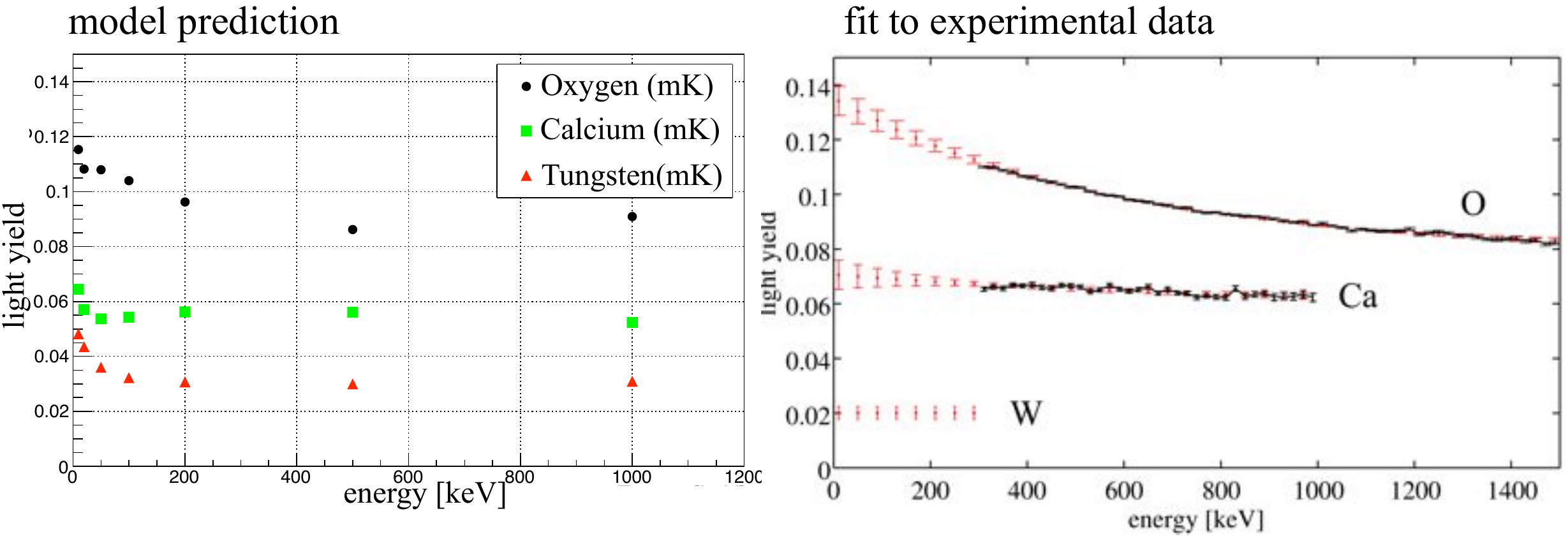}
  \caption{Left: Nuclear recoil light-yield for O, Ca and W recoils (10keV to 1MeV) induced, e.g., by neutrons or dark matter particles at mK temperatures as calculated with the microscopic model for the investigated CRESST crystal named \textit{Olga}. Right: Fit results of the light-yield measurement of neutron-induced O, Ca and W nuclear-recoils with a dedicated cryogenic detector module. Experiments carried out at the accelerator of the Maier-Leibnitz-Laboratory \cite{StraussPhD, Strauss}. Figure taken from \cite{StraussPhD}. Both figures show that the light yield depends on both, the type of the particle and its energy. The energy dependence is most pronounced for O recoils.}\label{LYnuclearrecoilsmodel}
 \end{center}
\end{figure*}
In figure \ref{ExpDTScold500nm} it can be seen that the experimentally determined decay-time spectra do follow the model description of an excitation-mode-dependent light-pulse shape. The model parameters were determined by fits of the light-pulse shape for the different excitation modes: \textit{unquenched} for rarely distributed STEs excited by laser excitation and \textit{quenched} for the excitation by iodine ions. For the excitation with oxygen ions the resulting prediction of the model for the light-pulse shape is shown: Good agreement between data and prediction is found.\\
\newline
\noindent A pertinent example for the application of the model is shown in figure \ref{LYnuclearrecoilsmodel}: The energy dependence of the nuclear recoil light yield as predicted by the model (left figure) and as determined experimentally (right figure, adopted from \cite{StraussPhD}). The light yield of an event caused by a certain particle "$part$" is defined as:
\begin{align}
LY^{part}_{det}(E_{det}) & = \frac{L^{part}_{det}(E_{det})}{E_{det}} 
\end{align}
where $L^{part}_{det}(E_{det})$ is the amount of scintillation light detected for the interaction of a particle "$part$" for which the energy $E_{det}$ is recorded in the CaWO$_4$ crystal.

\noindent In both figures, a particle-dependent energy dependence of the light yields can be seen. This energy dependence can be understood as a consequence of the energy (velocity) dependence of the ionization densities produced by the interacting particle: For energies in the keV regime, as considered here, it holds true that the smaller the primary ion velocity, the smaller the fraction of its energy used to produce ionization as well as the smaller the electronic stopping power. Thus, the smaller the ion velocity, the smaller the produced ionization density. Due to the strong dependence of the quenching effect on the ionization density, a less severe quenching effect is expected for less energetic ions, leading to an increased light yield compared to higher energetic ions. In addition, it should be noted that the light yield not only depends on the velocity of the interacting particle but additionally on its type: This is included via the particle-type dependent spatial model (see figure \ref{GeometricalModelIonizNucleus}) and the simulation of the produced ionization distribution. Thus, the model comprises the following dependencies: the particle-type and energy-dependent ionization production and its spatial distribution, the crystal- and temperature-dependent efficiency and temporal evolution of the STE-production, the crystal- and temperature-dependent STE deexcitation, i.e., the efficiency and temporal evolution of the scintillation light production and its extraction from the crystal.\\
\noindent The small differences of the light yields predicted by the model and determined experimentally (see figure \ref{LYnuclearrecoilsmodel}, deviations of $\sim 10$-$15\%$) can be explained by a combination of several effects: The usage of different crystals (with different optical quality) for the determination of the model parameters as well as for the experimental measurements \cite{StraussPhD, Strauss}; the current uncertainty of the model-parameter values due to the up-to-now limited time resolution ($\sim 20$ns) of the measurements utilized for the parameter determination; the uncertainty in the calculated ionization-density distribution of very low energy ions ($E_{ion}\lesssim20$keV). \\
\noindent In order to overcome these deficiencies and reduce the model uncertainties in the future, we are currently setting up an upgraded experiment with greatly enhanced time resolution of the excitation source as well as of the light detection and data acquisition (targeted resolution: $0.5$ns), and with an increased laser power. With this table-top set-up, excitation of the investigated CaWO$_4$ crystal with the 2-photon effect (leading to rare as well as to dense excitations, depending on the focussing of the laser beam) as well as with electrons (leading to dense, particle-induced excitations) will be possible. In this way, a determination of the model parameters with a greatly improved precision will be reached.\\
\newline 
%============================================================================================================
In summary, we have presented a comprehensive theoretical model for the scintillation-light production and quenching in CaWO$_4$ crystals. For the first time, this model allows the calculation (on a microscopic basis) of the pulse shape and the amount of scintillation light produced by an interacting particle dependent on its type and energy as well as on the temperature and defect density of the regarded crystal. These dependences of the light yield can consistently be explained within the model by the dependence of the strength of the quenching effect on the ionization-density produced in the energy-loss process of the interacting particle. As discussed in detail in \cite{RothPhD}, also the scintillator non-proportionality (energy dependence of the electron-light yield) as well as the $\gamma$-quenching \footnote{A slight reduction of the $\gamma$-particle light yield in comparison to equally energetic electron-induced events.} (see e.g., \cite{Lang}) can be explained by a particle- and energy-dependent ionization density, i.e. by a quenching effect.\\
\noindent To transfer this model to other scintillators such as different inorganic crystal scintillators, liquid or gaseous nobel gases or organic liquid scintillators, comparably simple experiments have to be carried out to determine the respective model parameters and (if necessary) possible extensions of the model, e.g., singlet and triplet radiative decays in liquid Ar. Table-top experiments with the currently improved set-up can be expected to suffice for a determination of all model parameters and to explore the potential of various scintillators in terms of their scintillation properties, particle-identification, and background-suppression characteristics.\\
%============================================================================================================
\newline
This research was supported by the DFG cluster of excellence 'Origin and Structure of the Universe', the Helmholtz Alliance for Astroparticle Physics, and the Maier-Leibnitz-Laboratorium (Garching). We like to thank the crystal laboratory of the Technische Universit\"at in M\"unchen for providing the crystal samples.

\end{document}